\documentclass[crystals,article,submit,moreauthors, pdftex,10pt,a4paper]{mdpi} 
\firstpage{1} 
\articlenumber{x}
\doinum{10.3390/------}
\pubvolume{xx}
\pubyear{2016}
\copyrightyear{2016}
\externaleditor{Academic Editor: name}
\history{Received: date; Accepted: date; Published: date}
\pdfoutput=1

 \theoremstyle{mdpi}
 \newcounter{thm}
 \newcounter{ex}
 \newcounter{re}
 \theoremstyle{mdpidefinition}

\Title{Quantum simulation of a 2D quasicrystal with cold atoms}

\Author{Nicolas Macé $^{1}$, Anuradha Jagannathan $^{1*}$ and Michel Duneau $^{2}$}
\AuthorNames{Nicolas Macé, Anuradha Jagannathan and Michel Duneau}

\address{%
$^{1}$ \quad Laboratoire de physique des Solides, Université Paris-Sud, 91405 Orsay, France\\
$^{2}$ \quad Centre de Physique Th\'{e}orique, \'{E}cole Polytechnique, Palaiseau, France}

\corres{Correspondence: jagannathan@lps.u-psud.fr; Tel.: +33-1-69156943}

\abstract{ We describe a way to obtain a two-dimensional quasiperiodic tiling with eight-fold symmetry using cold atoms. One can obtain a series of such optical tilings, related by scale transformations, for a series of specific values of the chemical potential of the atoms. A theoretical model for the optical system is described and compared with that of the well-known cut-and-project method for the Ammann-Beenker tiling. The relation between the two tilings is discussed. This type of cold atom structure should allow the simulation of several important lattice models for interacting quantum particles and spins in quasicrystals. }


\begin{document}



\section{Introduction}

With the discovery of the first quasicrystals \cite{schecht}, the quest began for, on the one hand, new quasiperiodic systems with better characterization of structural properties, and on the other hand, for theoretical methods to handle these systems. One of the goals of experiment has been, in particular, obtaining a single component quasicrystal, in the hope of finding direct relationships between its physical and geometrical properties. This may, we hope, become possible in cold atom systems. Cold atoms in optical lattices have been used to simulate quantum behavior of periodic crystals but not, thus far, of quasiperiodic tilings.   Cold gases, of cesium, rubidium or potassium atoms for example,  are used as  quantum simulators for a great variety of systems. As compared to real solid state systems, cold atom systems represent ideal systems, in which model parameters can be tuned at will.  Improvements in experimental techniques has resulted in an explosion of experimental simulations of condensed matter physics, quantum information and quantum optics models. Ultracold quantum gases in optical potentials therefore provide an exciting possibility towards the goal of synthesizing a perfect one-component quasicrystal.

The behavior of electrons in quasicrystals remains insufficiently understood, especially in dimensions higher than one.  Numerical investigations have thus far been limited by the amount of computational time needed. Much effort has been devoted to trying to understand tight-binding models on simple tilings, as a first step towards the description of more realistic systems. It is thus natural to ask what possibilities exist for realizing a quasiperiodic tiling by trapping cold atoms in an optical potential. The advantages of such a system, if realized, are manifold. First, it would become possible to fabricate samples using a single atomic species -- a significant simplification compared to real quasicrystals. It would be possible to directly observe the quantum states of the atoms, described by a tight-binding Hamiltonian. Finally and importantly, many of the parameters of the hopping and interaction Hamiltonians could be tuned. Many body properties in quasiperiodic systems could be properly studied under controlled conditions. The amount of disorder could be tuned as desired, to mimic experimental situations.
An experimental set-up to realize a two dimensional tiling was proposed in \cite{europhys2013,epjb2014}. It was shown that one can obtain an eight-fold tiling bearing a close relation to the Ammann-Beenker tiling \cite{beenker}. We will present the experimental system, the theoretical model and explain how a perfect quasiperiodic tiling can be obtained by introducing additional small interactions between atoms. 

This paper begins with a description of the experimental set-up. We then introduce a 4D description of the optical tiling which is naturally suggested by the experimental geometry. We next show the relation between the optical tilings and the perfect Ammann Beenker tiling, and describe how to transform the former into the other. In conclusion, some directions for simulating important theoretical models are suggested.

\section{Experimental set-up}
Optical trapping of atoms by laser generated potentials has led to the artificial realization in experiments of many different kinds of models on lattices, in which particles can move and interact via experimentally controlled interactions \cite{bloch}.  
Atoms can be trapped by laser light thanks to the dipole force acting on an atom due to the Stark effect in an off-resonance electric field. Under suitable assumptions concerning the decay rate $\Gamma$ of the excited state (for more details see the review ), it can be shown that the potential energy landscape seen by the atom has the form

$$ V(\vec{r}) =  v_0 I(\vec{r}) $$

where $v_0$ is a constant and $I(\vec{r})$ is the average value of the intensity at the position $\vec{r}$. Denoting the electric field by $\vec{E}$,  $I(\vec{r})=\vert\vec{E}(\vec{r}) \vert^2$. The sign of the prefactor, $v_0$, depends on the polarizability of the atom, and this can be positive or negative depending on the frequency of the laser, $\omega_L$, relative to the resonance frequency $\omega_0$ for the atom \cite{grimm2000}. The constant $v_0$ depends, among other parameters, on the laser detuning and is positive for blue-detuning ($\omega_L>\omega_0$) and negative for red-detuning ($\omega_L <\omega_0$). In other words, atoms experience a net force directed towards nodes (antinodes) of the laser intensity pattern if the detuning parameter $\Delta=(\omega_L-\omega_0)$ is positive (resp. negative).  In this paper we study this latter case, corresponding to atoms being attracted to maxima of the laser intensity pattern $I$. We note that, for large detuning, there will appear corrections to this interaction potential due to non-conservative processes and we will neglect these. Using this type of optical potential it has been possible to simulate models defined on a large variety of periodic structures including the square, triangular, kagome and other lattices. We will now consider a quasiperiodic structure with eight-fold symmetry obtained by this method.

\begin{figure}[!ht]
\centering
\includegraphics[width=120pt]{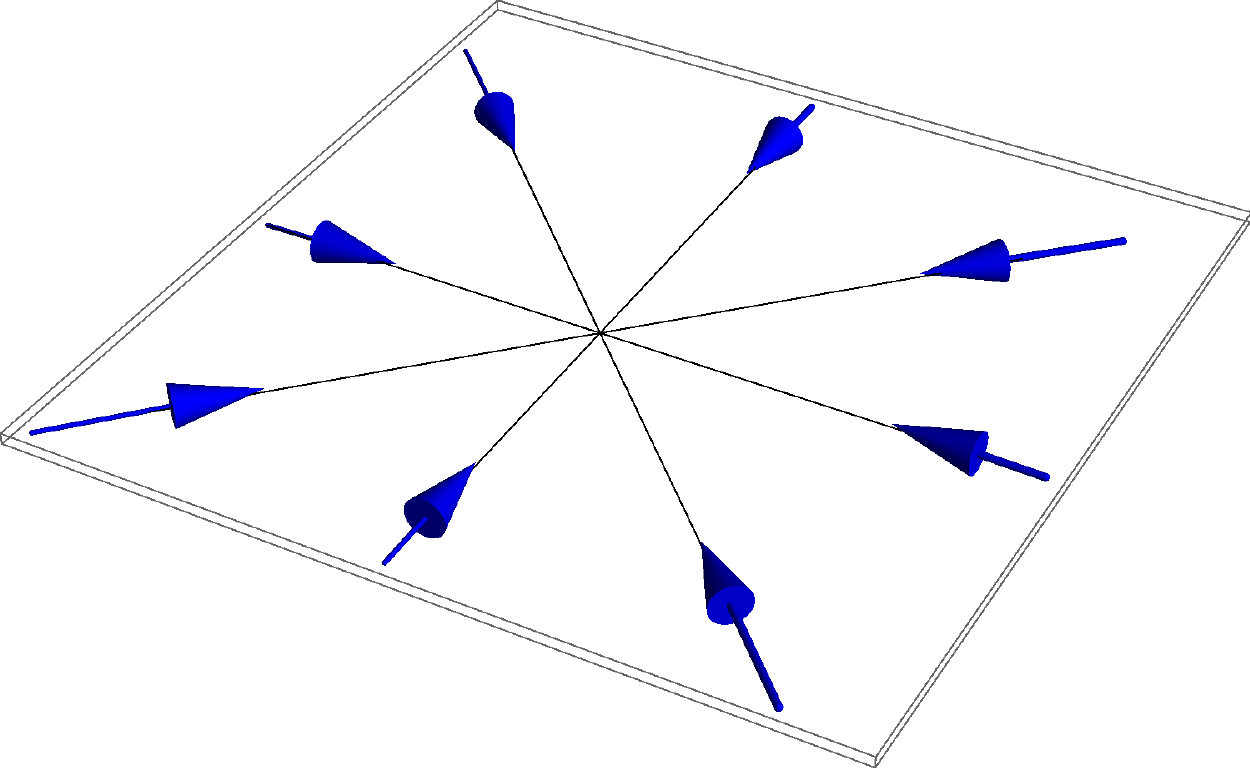}
\caption{Schematic of experimental set-up with four standing waves at $\pi/4$ angles. The light is polarized perpendicularly to the plane of propagation.}
\label{laser.fig}
\end{figure}

We consider a region where standing waves have been set up using four laser beams oriented at 45$^\circ$ angles in the $xy$ plane, as shown in Fig.\ref{laser.fig}. The wavelengths, $\lambda$, are the same for all the beams, as are their amplitudes. We will consider the situation where all the polarizations are perpendicular to this plane, allowing the amplitudes to sum up to an absolute maximum. The alternative choice, of using in-plane polarizations yields smaller maxima of amplitudes, smaller contrast, and would therefore be less efficient in trapping particles.
For the case of four standing waves,  the intensity is given by
\begin{equation}
I(\vec{r}) =  I_0 \left[\sum_{n=1}^4 \cos(\vec{k}_n.\vec{r}+\phi_n)\right]^2 
\label{Ixy.eq}
\end{equation}

where $\vec{r}=(x,y)$ is the position vector of the points lying in the plane of the lasers. The four wave vectors are given by
\begin{equation}
\vec{k}_n=k(\cos\theta_n,\sin\theta_n) \qquad \qquad \theta_n=\frac{(n-1)\pi}{4}
\label{kvecs.eq}
\end{equation}
with $n=1,..,4$, where $k=2\pi/\lambda$. Notice that the four beams can have arbitrary different phase-shifts $\phi_n$. As long as the relative phase shifts are maintained at some fixed arbitrary values, these phase shifts do not change the nature of the structures obtained, as discussed later. 

The function $I(\vec{r})$ of Eq.\ref{Ixy.eq} is quasiperiodic since there is no integer relationship between the four wave vectors $\vec{k}_n$. The intensity landscape obtained for a random choice of the phases $\phi_n$ has a complex structure of maxima, minima and saddlepoints. The intensities of the peaks, or local maxima, have a range of values with the upper bound $I_{max}=16I_0$. The potential energy thus has a minimum value of $V_{min}=-16V_0$ with $V_0=\vert v_0\vert I_0$.

The cold atoms in this region are attracted to local maxima of $I(r)$ - i.e. the local minima of the potential energy. If one now introduces a finite density of atoms, depending on the chemical potential, only maxima corresponding to intensities bigger than a cut-off $I_c$ will be occupied by an atom. We will neglect fluctuations due to finite temperature and trapping of atoms in metastable configurations, and instead focus on the ideal structures one expects to find, as a function of $I_c$.

\begin{figure}[!ht]
\centering
\includegraphics[width=250pt]{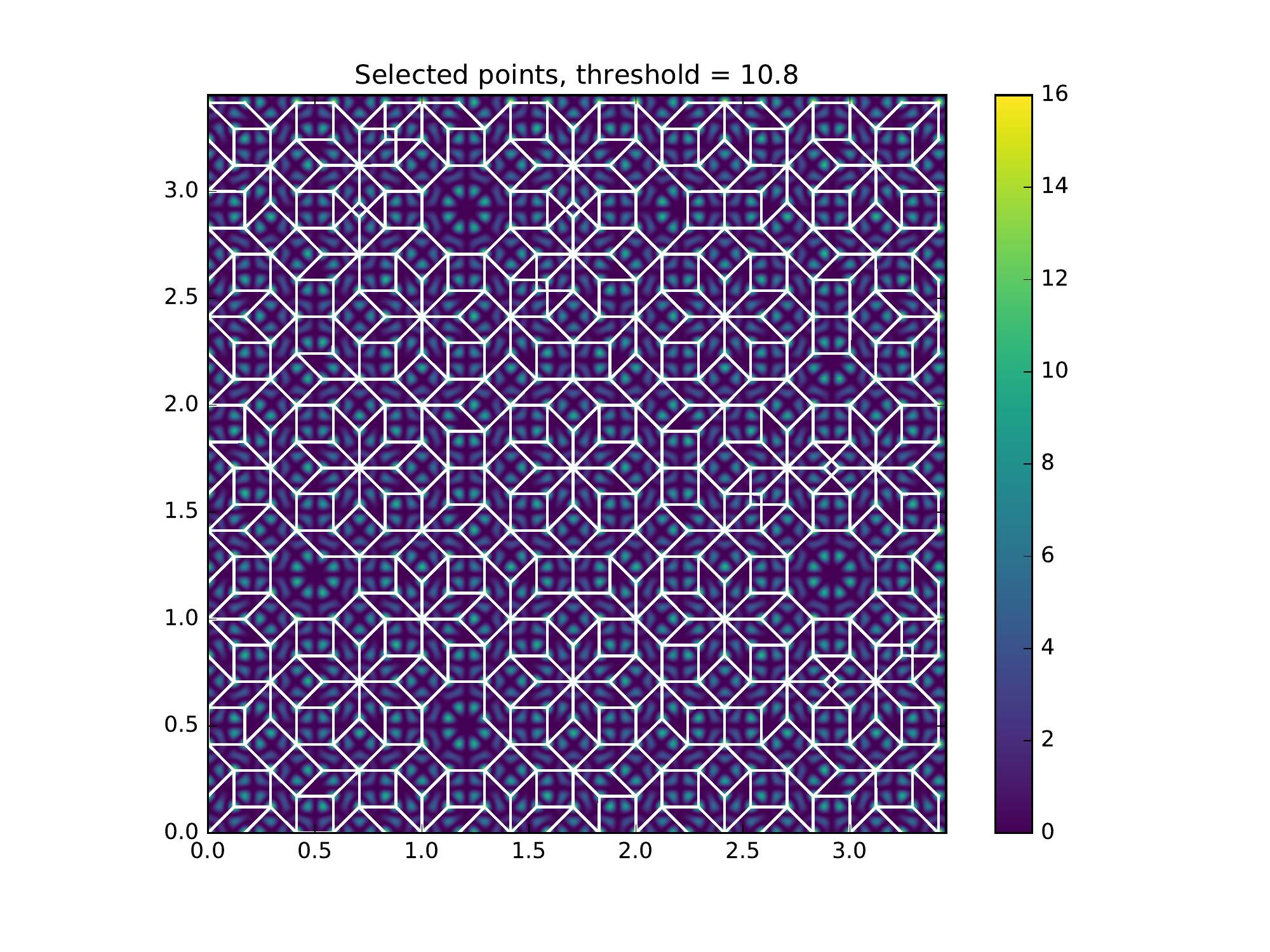}
\includegraphics[width=250pt]{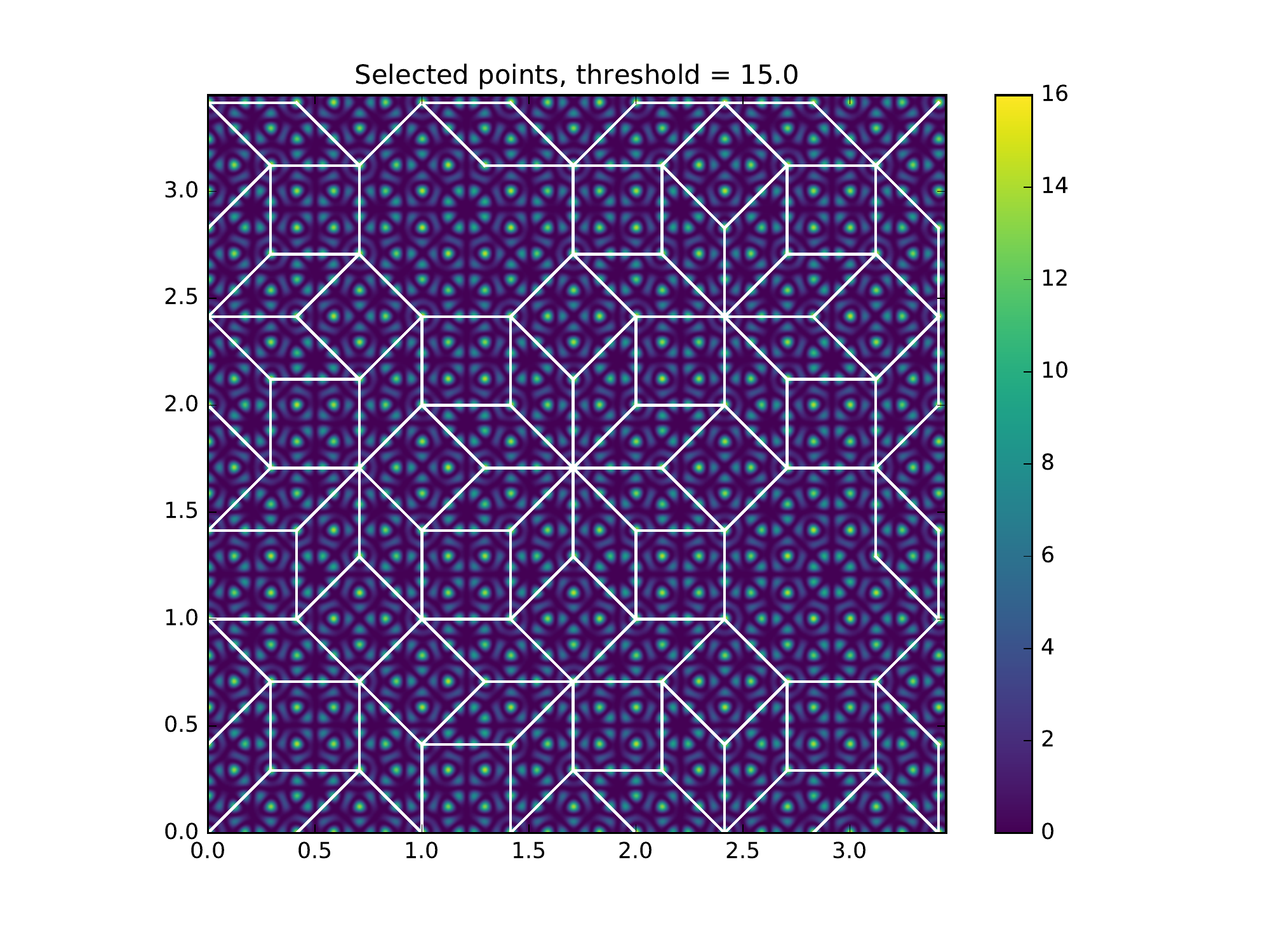}
\includegraphics[width=250pt]{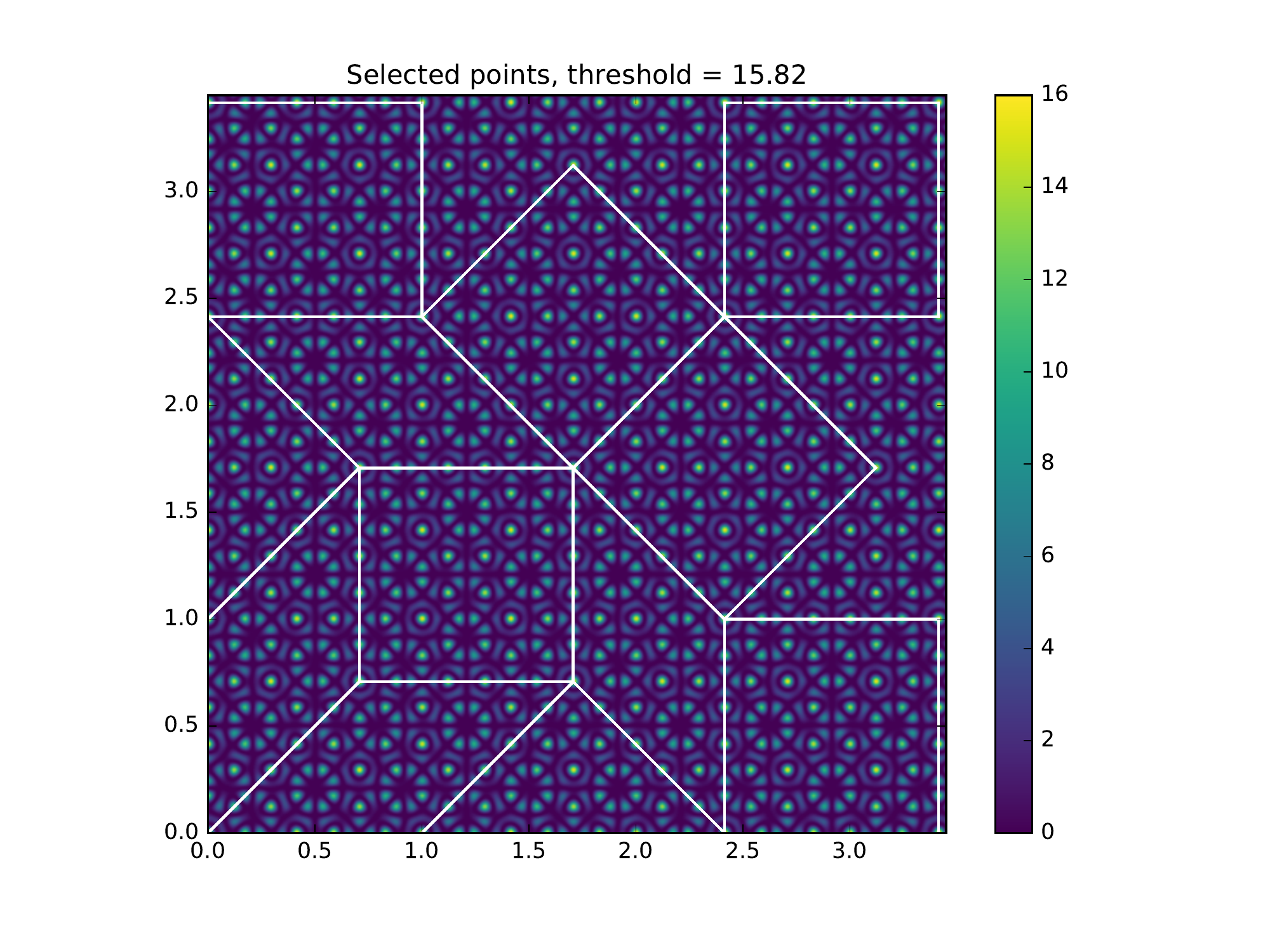}
\caption{Intensity plot showing the profile of the optical potential $I(\vec{r})$ in the plane. On this are superposed the tilings obtained by connecting occupied sites, for intensity cut-offs $I_c/I_0= 10.8, 15.0$ and $15.82$.   }
\label{twotilings.fig}
\end{figure}

Figs.\ref{twotilings.fig} show the optical tilings obtained for three representative values of the intensity cut-off: $I_c/I_0=10.8, 15$  and $15.82$. The edge-length of the tiles can be seen to increase by a discrete scale factor, the irrational number $ \alpha=1+\sqrt{2}$,  also called the silver mean. The tilings are shown superposed on the intensity profile, represented by a shaded plot (dark shades for small intensity). These are examples of the type of structure that we will refer to as the optical quasicrystal (OT). As can be seen from the shape of the tiles, it is closely related to the standard Ammann-Beenker or octagonal tiling (ABT) \cite{beenker,octagonal2} composed of squares and 45$^\circ$ rhombuses. For larger and larger values of the cutoff, approaching $16 I_0$, only the largest maxima will be occupied, and the  atomic density in the $xy$ plane correspondingly decreases. As the cutoff takes on successive special values in the series given in the next section, the lattice vectors increase by powers of the irrational number $\alpha$.

\section{A four dimensional model for the optical quasicrystal}
This section summarizes the model of the OT which was introduced in \cite{europhys2013,epjb2014}, along with some additional details, and clarifications.  Our aim here will be to give a mathematical description of the optical tilings shown in Fig.\ref{twotilings.fig}, and calculate quantities such as the lattice vectors in terms of the cutoff intensity $I_c$. In this model, the only dimensionful parameters are $\lambda=2\pi/k$ the wavelength of the laser light, which sets the length-scale, and $V_0$, which sets the energy scale in the problem. All other relevant lengths and energies can then be specified in terms of these two quantities.

\begin{figure}[!ht]
\centering
\includegraphics[width=260pt]{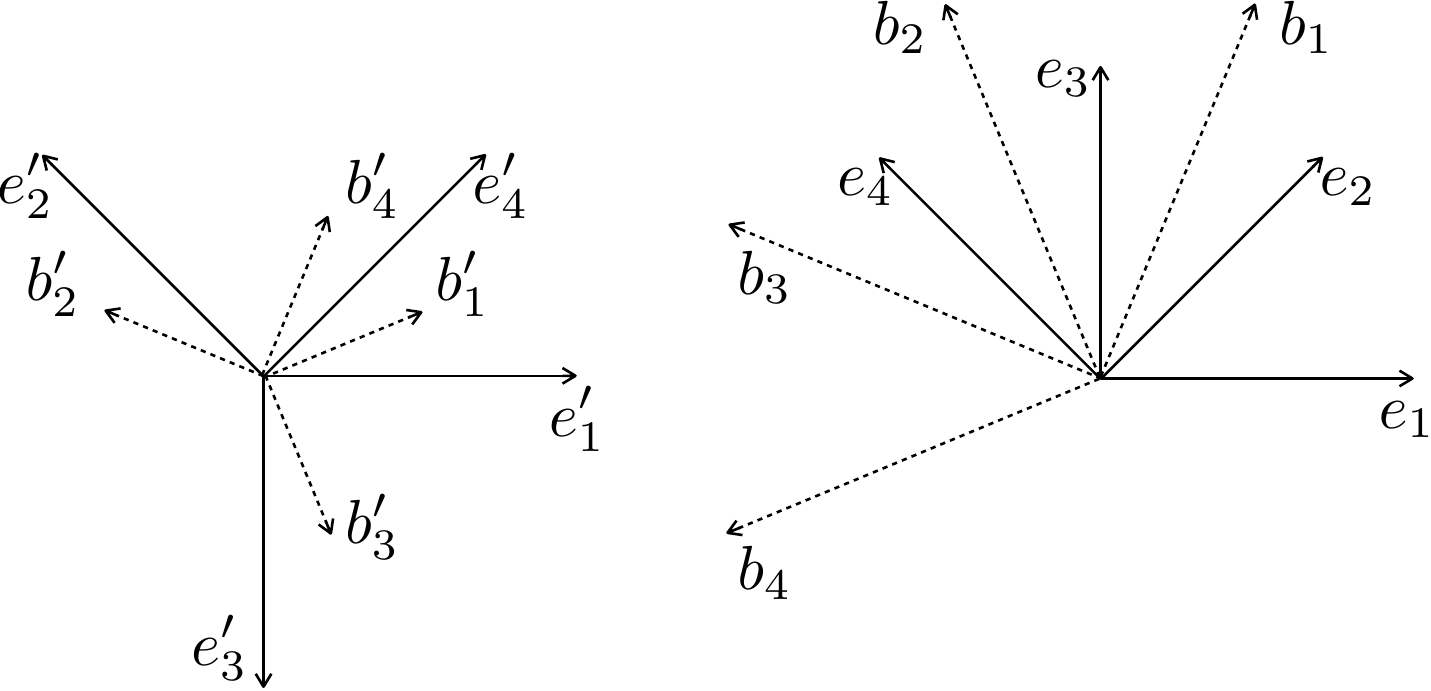}
\caption { Projections of the bases $\{\varepsilon_n\}$ of $\mathbb{Z}^4$ and $\{\beta_n\}$ of 
of the BCC lattice in $P'$ (left) and $P$ (right).}
\label{vecs.fig}
\end{figure}

Let us return to the intensity of the laser beams given by the function Eq.\ref{Ixy.eq}.
Since the four vectors $\vec{k}_n$ have no rational relationships $I(\vec{r}) $ is a quasiperiodic 
function in the sense of H. Bohr \cite{bohr} and A.S. Besicovitch \cite{besic}. 
The Fourier transform is readily obtained by expanding 
cosines so that the spectrum is the finite set of all combinations $\pm \vec{k}_n\pm \vec{k}_m$. \\
The 8-fold symmetry of $I(\vec{r})$ follows from the remark that a rotation $\gamma$ (see \cite{epjb2014} of $\pi/4$ 
maps $\{\vec{k}_1,\vec{k}_2,\vec{k}_3,\vec{k}_4\}$ to $\{\vec{k}_2,\vec{k}_3,\vec{k}_4,-\vec{k}_1\}$ 
(see Fig. \ref{vecs.fig}). 
$I(\vec{r})$ is also invariant by the 2-fold symmetry $\sigma$ w.r.t. the $x$-axis, which 
maps $\{\vec{k}_1,\vec{k}_2,\vec{k}_3,\vec{k}_4\}$ to $\{\vec{k}_1,-\vec{k}_4,-\vec{k}_3,-\vec{k}_2\}$. 
While this $C_{8v}$ symmetry is not crystallographic in 2D, it is crystallographic in the 
4D space $\mathbb{R}^4$: Lifting $\gamma$ and $\sigma$ to $\mathbb{R}^4$ gives the integer matrices 
\begin{align}
\Gamma =  \left[\begin{array}{rrrr}
0 & 0 & 0 & -1 \\
1 &  0 & 0 & 0 \\
0 & \phantom{-}1 & 0 & 0 \\
0 & 0 & \phantom{-}1 & 0
\end{array}\right], \quad
\Sigma = \left[\begin{array}{rrrr}
1 & 0 & 0 & 0 \\
0 & 0 & 0 & -1 \\
0 & 0 & -1 & 0 \\
0 & -1 & 0 & 0
\end{array}\right],
\end{align}
in the standard basis $\{\vec{\varepsilon}_1,\vec{\varepsilon}_2,\vec{\varepsilon}
_3,\vec{\varepsilon}_4\}$. 
They satisfy $\Gamma^8=\Sigma^2=I$ and $\Sigma\Gamma\Sigma^{-1}=\Gamma^{-1}$. 
This 4D representation of $C_{8v}$ is reducible since $\mathbb{R}^4$ is the Cartesian product  
of two planes $P$ and $P'$ which are orthogonal and invariant. 
While the restriction of $\Gamma$ to $P$ is the previous $\gamma$ rotation, 
the restriction to $P'$ is a rotation $\gamma'$ of $3\pi/4$. \\
One can choose orthonormal bases $\{\vec{e}_x,\vec{e}_y\}$ in $P$ (the ''physical space" ), 
and $\{\vec{e}'_x,\vec{e}'_y\}$ in $P'$ (the ''perpendicular space") 
so that the orthogonal projections $\vec{e_n}=\boldsymbol{\pi}(\vec{\varepsilon}_n)$ and  
$\vec{e}'_n=\boldsymbol{\pi}'(\vec{\varepsilon}_n)$, all of norm $1/\sqrt{2}$, are as shown in 
Fig. \ref{vecs.fig}. \\
Points $\vec{R}=(R_1,R_2,R_3,R_4)$ of  $\mathbb{R}^4$ also write 
$(\vec{r},\vec{r}')=(x,y,x',y')$ in the $\{\vec{e}_x,\vec{e}_y,\vec{e}'_x,\vec{e}'_y\}$ bases of 
$\mathbb{R}^4$. The transformation is given by the following rotation $\mathcal{R}$: 
\begin{align*}
\left[\begin{array}{l}
x\\y\\x'\\y'
\end{array}\right] &=\mathcal{R}
\left[\begin{array}{l}
R_1\\R_2\\R_3\\R_4
\end{array}\right]
=\frac 1 2
\left[\begin{array}{rrrr}
\sqrt{2} & 1 & 0 & -1\\
0 & 1 & \sqrt{2} & 1 \\
\sqrt{2} & -1 & 0 & 1 \\
0 & 1 & -\sqrt{2} & 1
\end{array}\right].
\left[\begin{array}{l}
R_1\\R_2\\R_3\\R_4
\end{array}\right].
\end{align*}
If $\vec{R}=(\vec{r},\vec{r}')$ and $\vec{K}=(\vec{k},\vec{k}')$ are two 4D vectors then 
$\vec{K}.\vec{R}=\sum K_n R_n=\vec{k}.\vec{r}+\vec{k}'.\vec{r}'$.  

The wave vectors $\vec{k}_n$ are the projections in $P$ of orthogonal 4D vectors 
$\vec{K}_n=K\vec{\varepsilon}_n=(\vec{k}_n,\vec{k}'_n)$, so that their magnitude 
is $K=\sqrt{2}\, k$. The intensity Eq. \ref{Ixy.eq} can be obtained from the 4D periodic function 
\begin{equation}
\mathcal{I}(\vec{R})=I_0 \left[\sum_{n=1}^4 \cos(\vec{K}_n.\vec{R})\right]^2, 
\label{V4d.eq}
\end{equation}
so that $I(\vec{r})=\mathcal{I}(\vec{r},0)$ is the restriction of $\mathcal{I}(\vec{R})$ to the plane $P$. 

The maxima of $\mathcal{I}(\vec{R})$ occur on points where all cosines equal $1$, i.e. on the 
cubic lattice $\frac{2\pi}{K}\mathbb{Z}^4$, and  on points where all cosines equal $-1$, which 
correspond to the body centers.
 Let $B$ denote the BCC lattice of $\mathbb{Z}^4$: coordinates are either all integers or all half 
 integers. Accordingly, the maxima $16\,I_0$ of $\mathcal{I}(\vec{R})$ are located on the BCC lattice 
$\frac{2\pi}{K}B=\frac{\sqrt{2}\pi}{k}B$. The vertices of $B$ can be written as 
$\vec{R}=T\vec{x}=\sum x_n\vec{\beta}_n$ with $\vec{x}=(x_1,x_2,x_3,x_4)$ 
(where the $x_n$ are integers) and where $T$ is the matrix 
\begin{align*}
T=\frac 1 2
\left[\begin{array}{rrrr}
1 & -1 & -1 & -1 \\
1 & 1 & -1 & -1 \\
1 & 1 & 1 & -1 \\
1 & 1 & 1 & 1
\end{array}\right].
\end{align*}
A basis of $B$ is given by the unit vectors $\vec{\beta}_n=T\vec{\varepsilon}_n$  ($n=1,..,4$). 
One can check that $\det(T)=\frac 1 2$ and that $T.C_{8v}=C_{8v}.T$ so that $B$ is invariant with respect to $C_{8v}$  and belongs to the same Bravais class as the hypercubic lattice $\mathbb{Z}^4$.
Furthermore, $T$ commutes with the 8-fold generator $\Gamma$. 
The four primitive lattice vectors $\vec{\beta}_n$ of $B$ project onto the set $\vec{b}_n$ in $P$, 
and $\vec{b}'_n$ in $P'$, as shown in Fig. \ref{vecs.fig}. The norms of these vectors are 
$b_n=\frac  1 2 \sqrt{2+\sqrt{2}}$ and $b'_n=\frac  1 2 \sqrt{2-\sqrt{2}}$. 


The condition $\mathcal{V}(\vec{R})\leq V_c$ is equivalent to a condition 
$\mathcal{I}(\vec{R})\geq I_c$. If the cutoff $I_c$ is high enough, the last condition is fulfilled in 
a set of disjoint domains centered on the BCC lattice $\frac{\sqrt{2}\pi}{k} B$. 
If  the cutoff $I_c$ is close to the absolute maximum $I_m=16\,I_0$, one can substitute the quadratic 
approximation  $\mathcal{I}(\vec{R})\approx (16-8k^2(\vec{r}^2+\vec{r}'^2))I_0$. The domains are 
close to spheres of radius $\rho$ given by $8k^2\rho^2=(16I_0-I_c)/I_0$. Their 
projections on $P$  and $P'$ are close to disks of the same radius. 

If $I(\vec{r})>I_c$ is a local maximum, the point $(\vec{r},\vec{0})$ belongs to a domain 
centered on a vertex $\vec{R}$ of the BCC lattice, at a distance bounded by $\rho$ in the quadratic 
approximation. These points $\vec{r}$ are close to the BCC lattice, allowing for a comparison with 
the cut-and-project method as discussed below.

\begin{figure}[!ht]
\centering
\includegraphics[width=200pt]{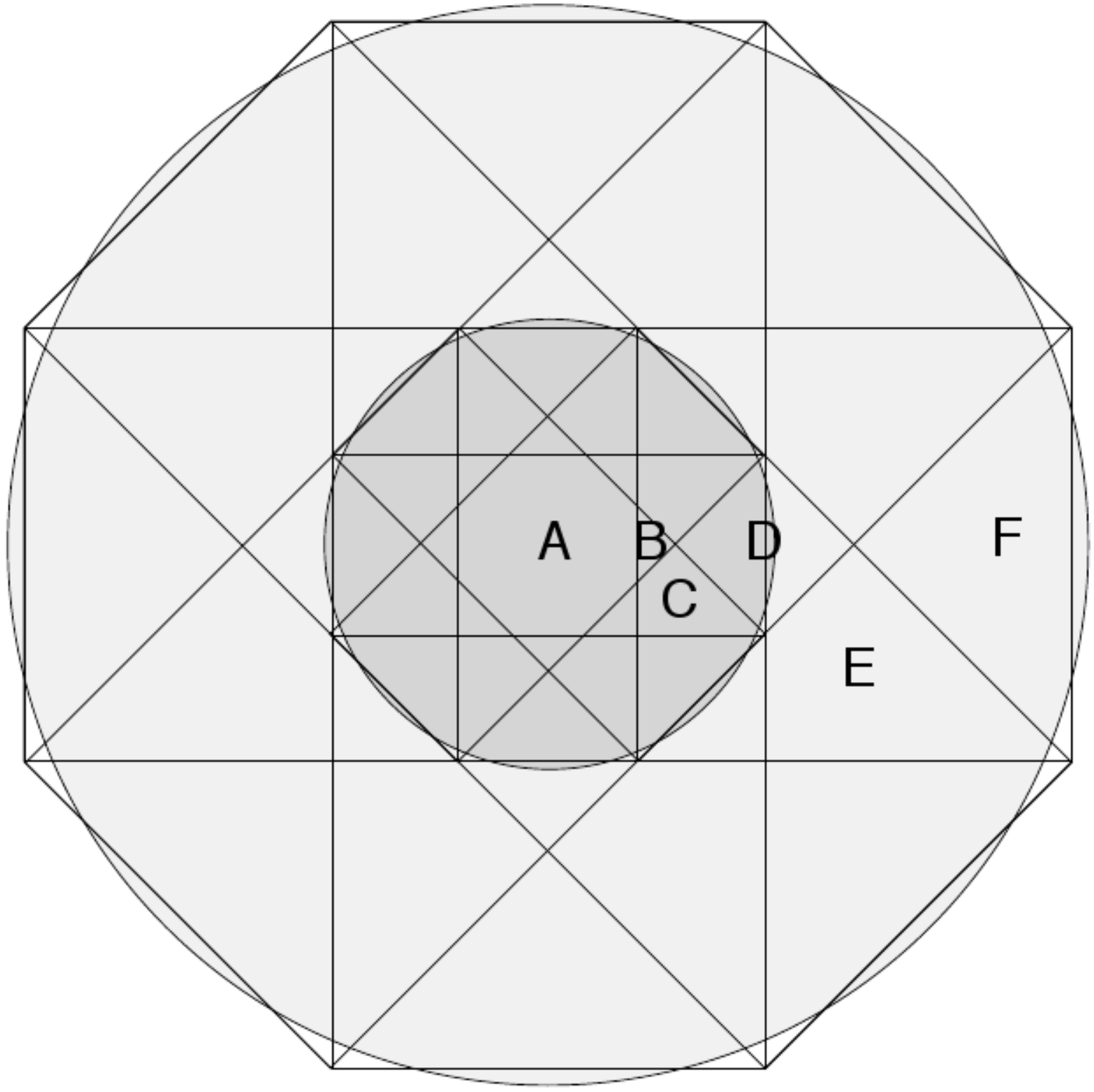}  \hskip 2cm
\includegraphics[width=120pt]{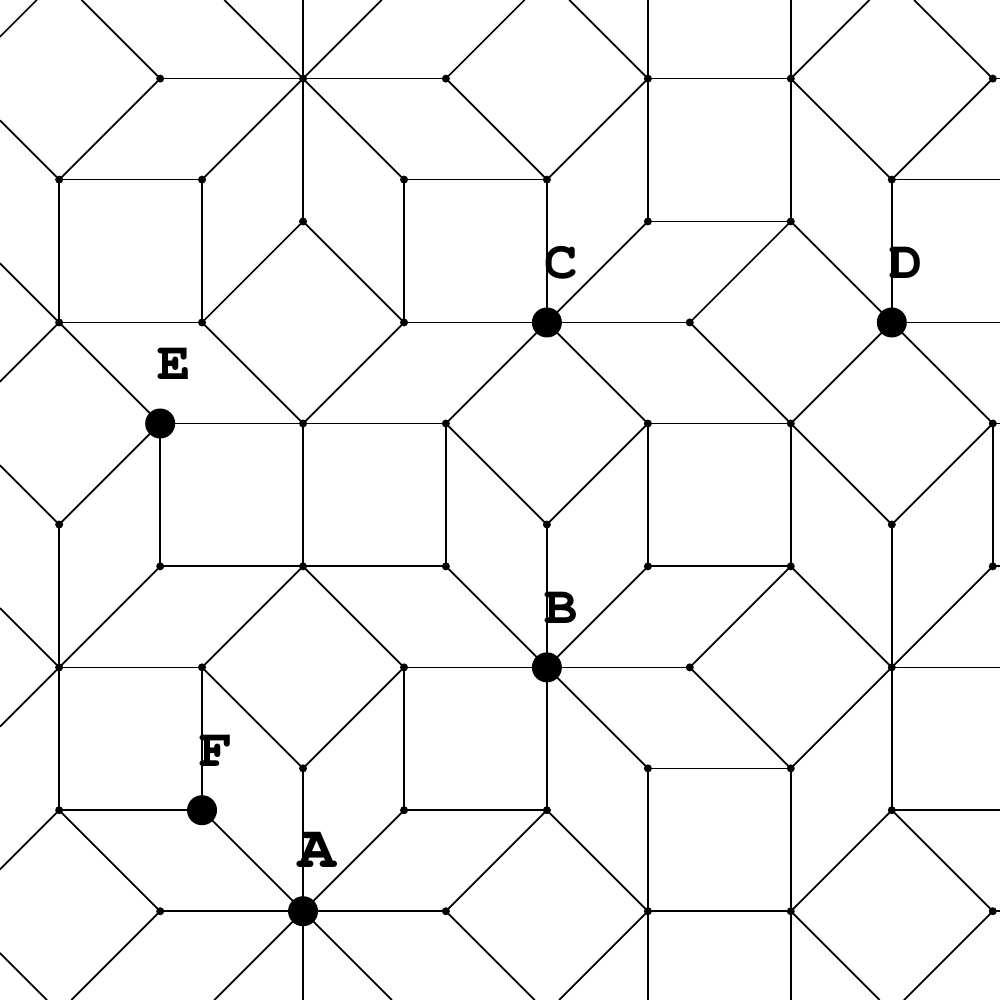} 
\caption{a) (left) The big octagon represents W, the selection window for the ABT in $P'$. 
Circles represent the selection windows $D'$ for the OT for the values $p=0$ and $p=1$. b). 
(right) A portion of the OT with the corresponding local environments (A,B,..) corresponding to subwindows of the big octagon.}
\label{windows.fig}
\end{figure}

The vertices of the octagonal tiling  \cite{octagonal2} are the projections $\vec{r}$ of 4D lattice points 
$\vec{R}=(\vec{r},\vec{r}')$ such that $\vec{r}'$ belongs to an octagonal window 
generated by the four vectors $\vec{b}'_n$ (see \cite{dunkatz} for the cut and project 
algorithm). 
The area of this window (see Fig.\ref{windows.fig}) is $W=\frac{\sqrt{2}\pi^2}{k^2}$. 
The inflation transformation of the octagonal tiling enlarges the edge length of the tiles in $P$  by a factor $\alpha$. while distances in $P'$ are reduced by the same factor.  Inflated octagonal tilings  correspond to selection windows of area $W/\alpha^{2p}$ with $p=0,1,2,..$.  

If $V_c$ is low enough and if the areas of $D'$  and $W$ are equal (up to inflation), a 
relationship between both structures can be expected. This equality ensures that  the two 
structures have the same density of points. 
Using the quadratic approximation of $\mathcal{I}(\vec{R})$ this condition writes
\begin{equation}
\frac{V_c-16\,V_0}{V_0}=\frac{16\,I_0-I_c}{I_0}=8\sqrt{2}\pi\alpha^{-2p}
\label{cond.eq}
\end{equation}

The OT of Fig.\ref{twotilings.fig} correspond to $p=1, 2$ and $3$.  The circular windows $D'$ are shown in Fig.\ref{windows.fig} for $p=0$ and $p=1$, inside the window of the octagonal tiling. The criterion in Eq.\ref{cond.eq} can be further simplified by noting that, for $p>1$, the cosine terms in $V$ can be expanded to second order in the distance $\rho$. In this limit the selection windows for the OT are approximately circular and the radius for the cutoff value $I_c$ is given by $k^2\rho_c^2=\sqrt{2}\pi\alpha^{-2p}$. It can be shown that the edges of the tiles of the $p$th OT have length $\ell=\frac{\sqrt{2}\pi}{k}\,|b_n|\alpha^p$.  The smallest edge length $\ell \approx 3.81 \lambda$ is obtained 
for $p=1$. 

\section{Fourier transform of the OT}

We now turn to the diffraction pattern of a structure obtained when atoms occupy all of the allowed vertices corresponding to a particular value of $I_c$. The general method for finding the Fourier transform and indexation of peaks is discussed for example by A. Katz and D. Gratias in \cite{diffrac}. The structure factor consists of Bragg peaks whose positions are given by the reciprocal lattice of the 4D cubic lattice, and whose peak intensities are given by the Fourier Transform of the selection window. Due to the close similarity of their selection windows, the structure factor of the OT and that of the ABT are expected to be very similar, with peaks in the same positions but with slight differences in their intensities.  Here we will confine our attention to the indexing of Bragg peaks.  It is interesting to observe that spherical selection windows were introduced by Grimm and Baake in the context of the diffraction patterns of quasiperiodic tilings \cite{grimmbaake}, as a useful approximation in the calculation of the structure factor. For the optical quasicrystal, in contrast, we see that the spherical window turns out to be the correct, physically imposed choice, provided only that the cut-off $I_c$ is large enough. \\

\begin{figure}[!ht]
\centering
\includegraphics[width=180pt]{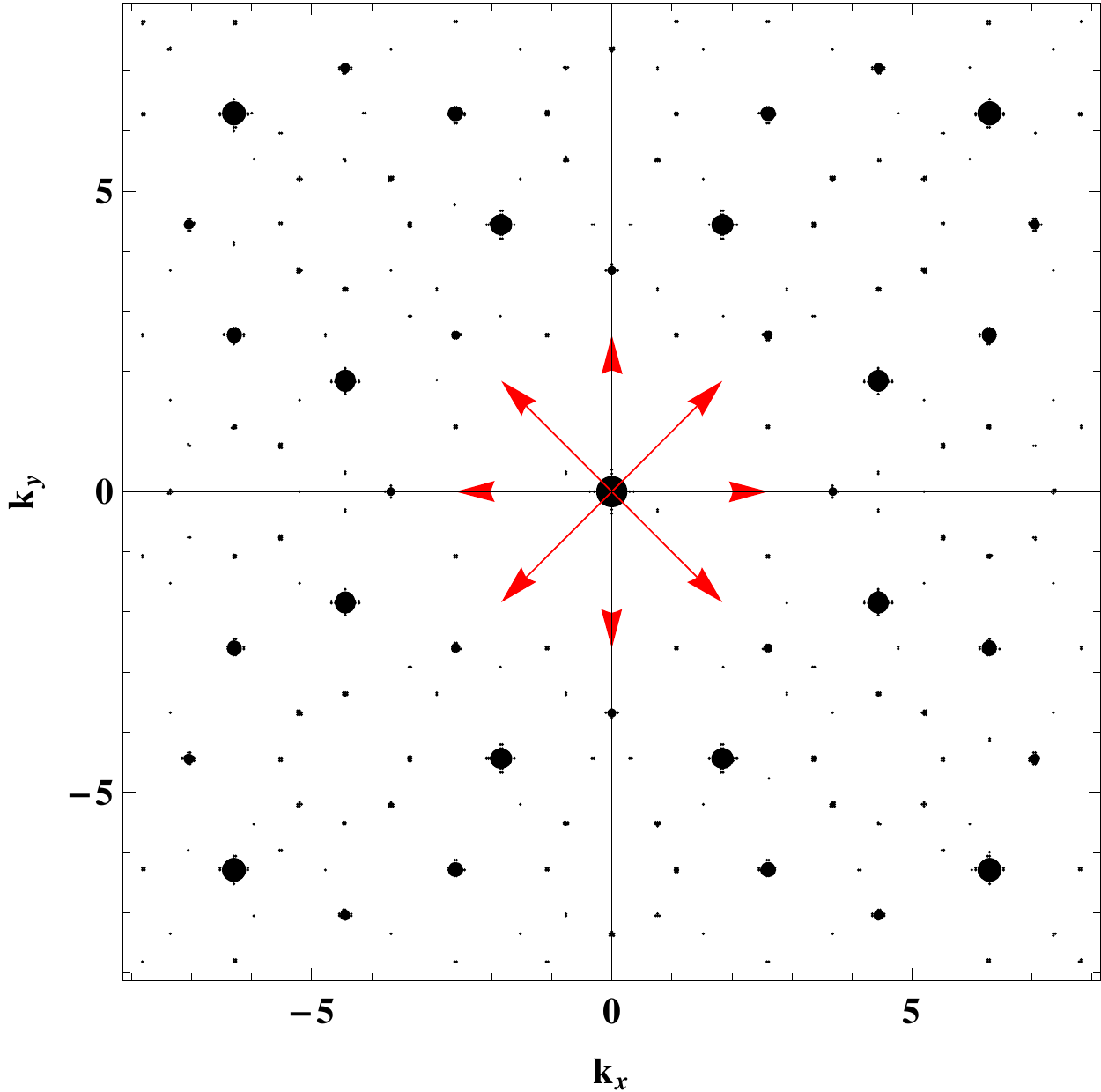}
\caption{ The structure factor calculated using Eq.\ref{struc.eq} for a sample of 4000 sites of the optical quasicrystal (p=0). Intensities of the peaks are proportional to the area of the spots, $k_x$ and $k_y$ are given in units of $\lambda^{-1}$. Arrows indicate the eight shortest reciprocal lattice vectors.}
\label{struc.fig}
\end{figure}

The 4D basis vectors are $\vec{K}_n = (2\sqrt{2} \pi/\lambda) \vec{\varepsilon}_n$. After projection in $P$, one obtains the four lattice vectors of Eq.\ref{kvecs.eq}
\begin{equation}
\vec{k}_n = \frac{2\sqrt{2}\pi}{\lambda}\vec{e}_n,
\end{equation}
 The peaks of the structure factor $S(\vec{k})$ are found at positions $ \vec{k}=\sum h_n \vec{k}_n$ with the condition that
\begin{equation}
\exp(i \pi \sum_n h_n) = 1
\label{hcond.eq}
\end{equation}
equivalent to the requirement that $\sum_n h_n$ be even.
This condition is equivalent to saying that the positions of its Bragg peaks are those of $k\sqrt{2}F$, the reciprocal lattice of $\frac{\pi\sqrt 2}{k}B$, after projection into $P$. The values of the intensities are determined by the Fourier transform of the selection window, which is of octagonal shape for the ABT, and circular shape for the OT. The resulting differences between the two structure factors would show up only under a detailed comparison of peak intensities. Fig.\ref{struc.fig} shows the eight vectors $\pm  \vec{k}_n$, and the numerically calculated intensity function $S(\vec{k})$ defined by

\begin{equation}
S(\vec{k}) = \frac{1}{N} \sum_{i,j} \exp[i \vec{k}.(\vec{r}_i-\vec{r}_j)]
\label{struc.eq}
\end{equation}
where $N$ is the number of sites in the sample. The set of intense peaks nearest the origin correspond to the combinations $\{\pm1,\pm1,0,0\}$ and permutations thereof, in accordance with the condition in Eq.\ref{hcond.eq}. The structure factors of successive OT for $p\geq1$ will be the same, only rescaled by powers of $ \alpha^{-p}$, to take into account the inflation in real space.

\section{From the optical tiling to the Ammann Beenker tiling}

\begin{figure}[!ht]
\centering
\includegraphics[width=200pt]{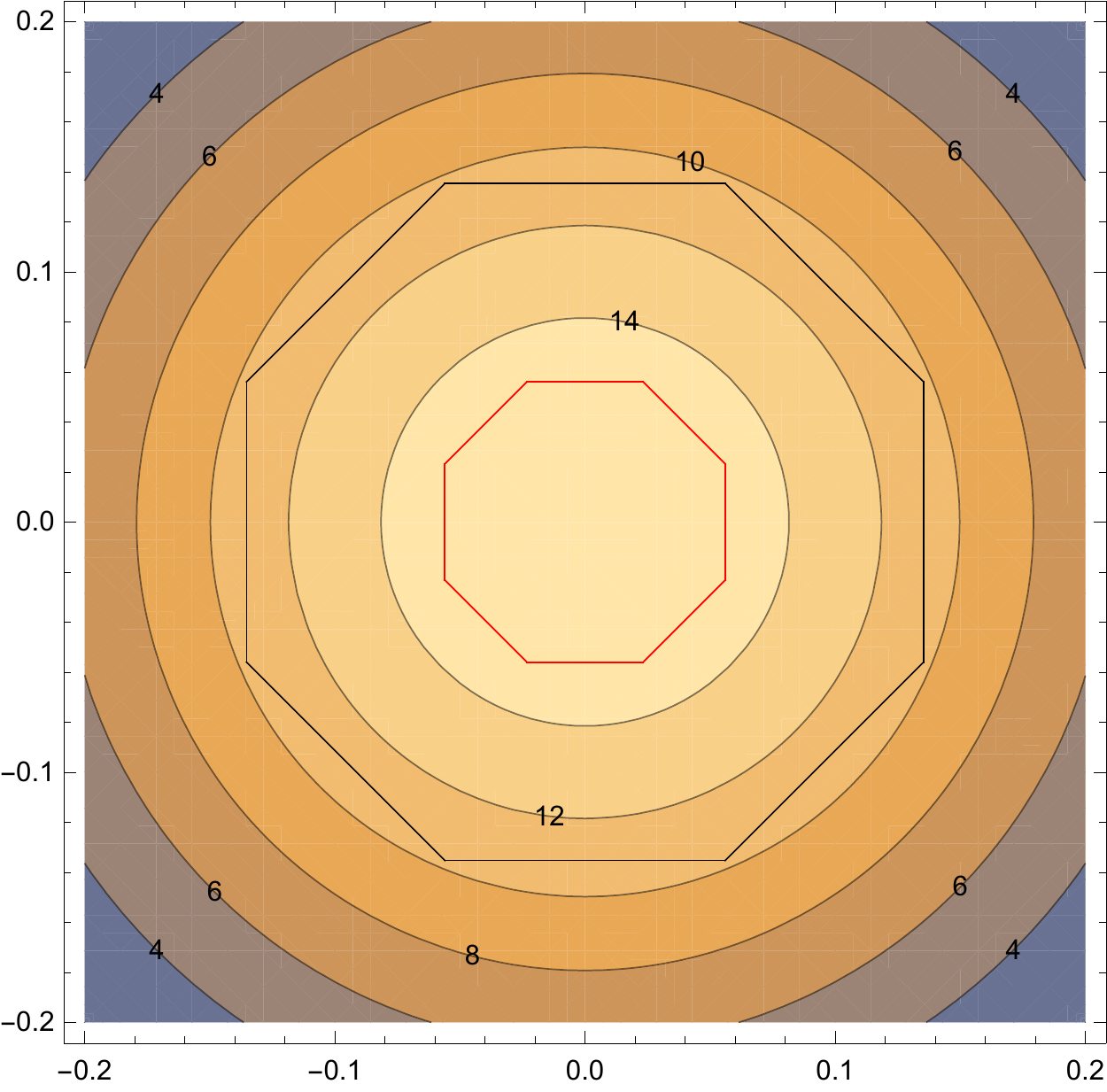} 
\caption{ Contour plot of the intensity of peaks as a function of the perpendicular space coordinates. The black octagon delimits the region corresponding to all sites of the p=1 OT,  and the red octagon delimits the region corresponding to $z\geq 5$ sites.}
\label{perppot.fig}
\end{figure}

In the preceding discussion, we assumed that when atoms are loaded into this optical potential, they occupy sites of lowest potential energy i.e., highest intensity, at very low temperature, giving rise to the optical tiling. While overall, the intensities of the occupied peaks vary within the range $I_{max} >I>I_c$,  one can define different categories of sites. It is easily seen that the typical value of the potential energy depends on the local environment, as follows. For convenience, we will describe the environment of each site by the letter A,B,... as determined by its position in the octagonal acceptance window W (see Fig.\ref{windows.fig}), and for purposes of illustration, we consider the OT for $p=1$. Fig.\ref{perppot.fig} is a contour plot of the intensity $I(\vec{r})$ in perpendicular space. The value of the potential on a given site depends on its perpendicular coordinate, decreasing with the distance from the origin. For the values of the potential shown, these contours are close to circular. On the other hand, as we saw  (in Fig.\ref{windows.fig}),  the distance from the origin in perpendicular space coordinate tends to get larger as $z$ decreases. As shown in Fig.\ref{perppot.fig}, the high coordination number sites ($8>z>5$) sites correspond to the region inside the small red octagon and these have the highest values  of intensity, $16>I/I_0>\approx 15$.  The sites of small coordination number, which lie in the region between the red and black octagons, have values in the range $15>\approx I/I_0>\approx  11$. 

\begin{figure}[!ht]
\centering
\includegraphics[width=180pt]{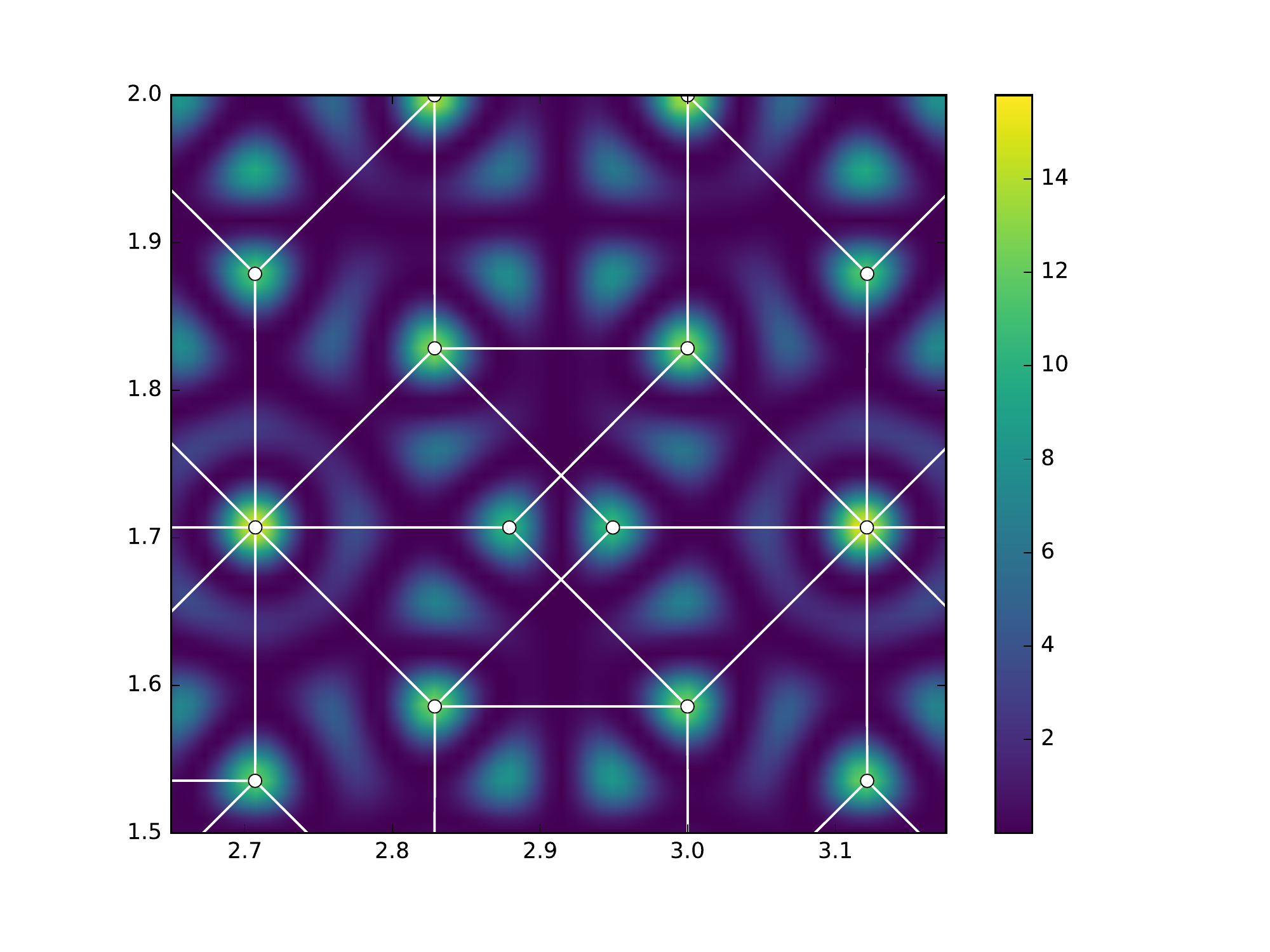} 
\includegraphics[width=180pt]{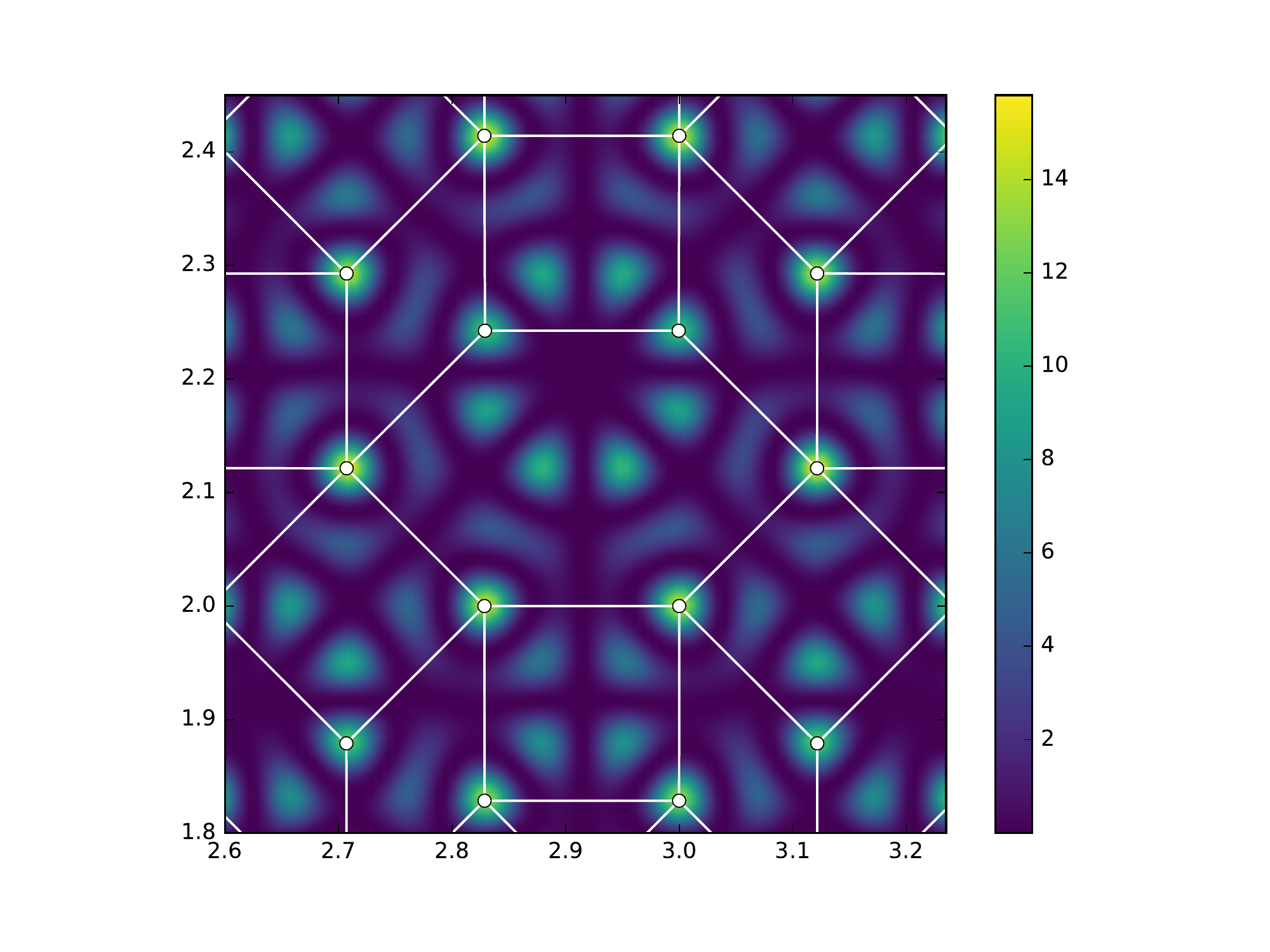} 
\caption{Two close-ups showing the intensity variation in the neighborhood a) of a pair of twin sites and b) of an empty hexagon of the optical quasicrystal}
\label{defects.fig}
\end{figure}

We turn now to the problem of structure differences between the Ammann Beenker tiling and the optical tiling. These arise from the difference in shape of their acceptance windows in perpendicular space. Comparing the two acceptance domains one sees that they overlap over most of the region they occupy. This ensures that a large fraction of the selected points are identical in the two tilings. The differences between the two windows arise in the outlying regions. The results in, firstly, some sites which are present in the AB tiling but are ``missing" in the OT. This leads to the empty hexagons and larger n-gons that one sees in Figs.2. Secondly, pairs of "twin-sites" separated by a very short distance $\delta$ appear in the OT, whereas in the AB tiling only one of the members of these pairs is present, the other being related by a ``phason-flip". These two types of differences are shown in detail Fig.\ref{defects.fig}. Differences of intensity of the occupied and unoccupied sites are too small to be detected visually. For example, in Fig.\ref{defects.fig}(a) the intensities of the twin sites are 10.8$V_0$ (left site) and 11.1$V_0$ (right site). Similarly,  in Fig.\ref{defects.fig}(b), the intensities of the bright spots in the center of the vacant hexagonal region are 10.5$V_0$ and 10.6$V_0$. \footnote{ The two defects shown are situated on the same vertical worm of the ABT.}

\begin{figure}[!ht]
\centering
\includegraphics[width=100pt]{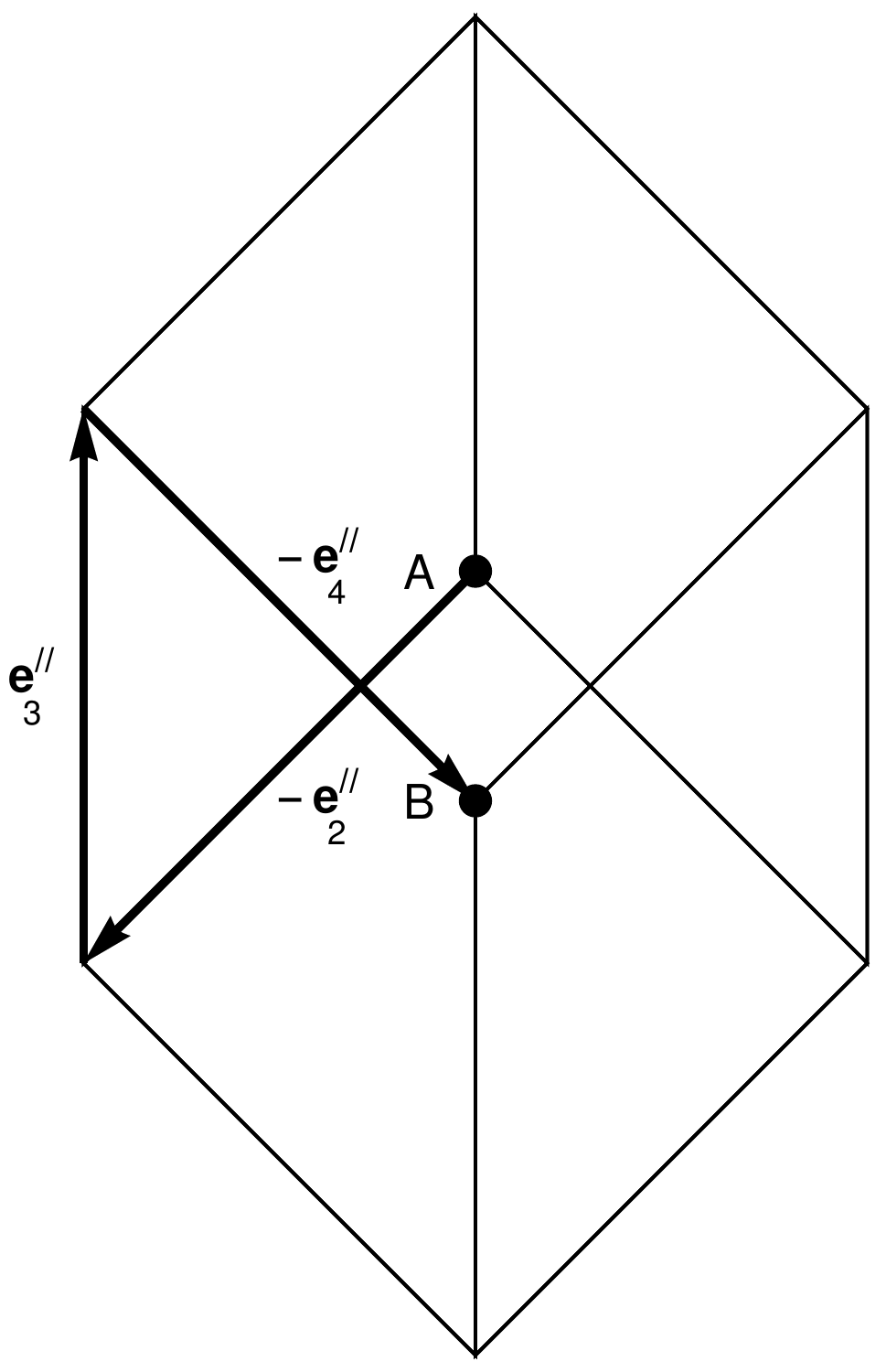} \hskip 2cm
\includegraphics[width=160pt]{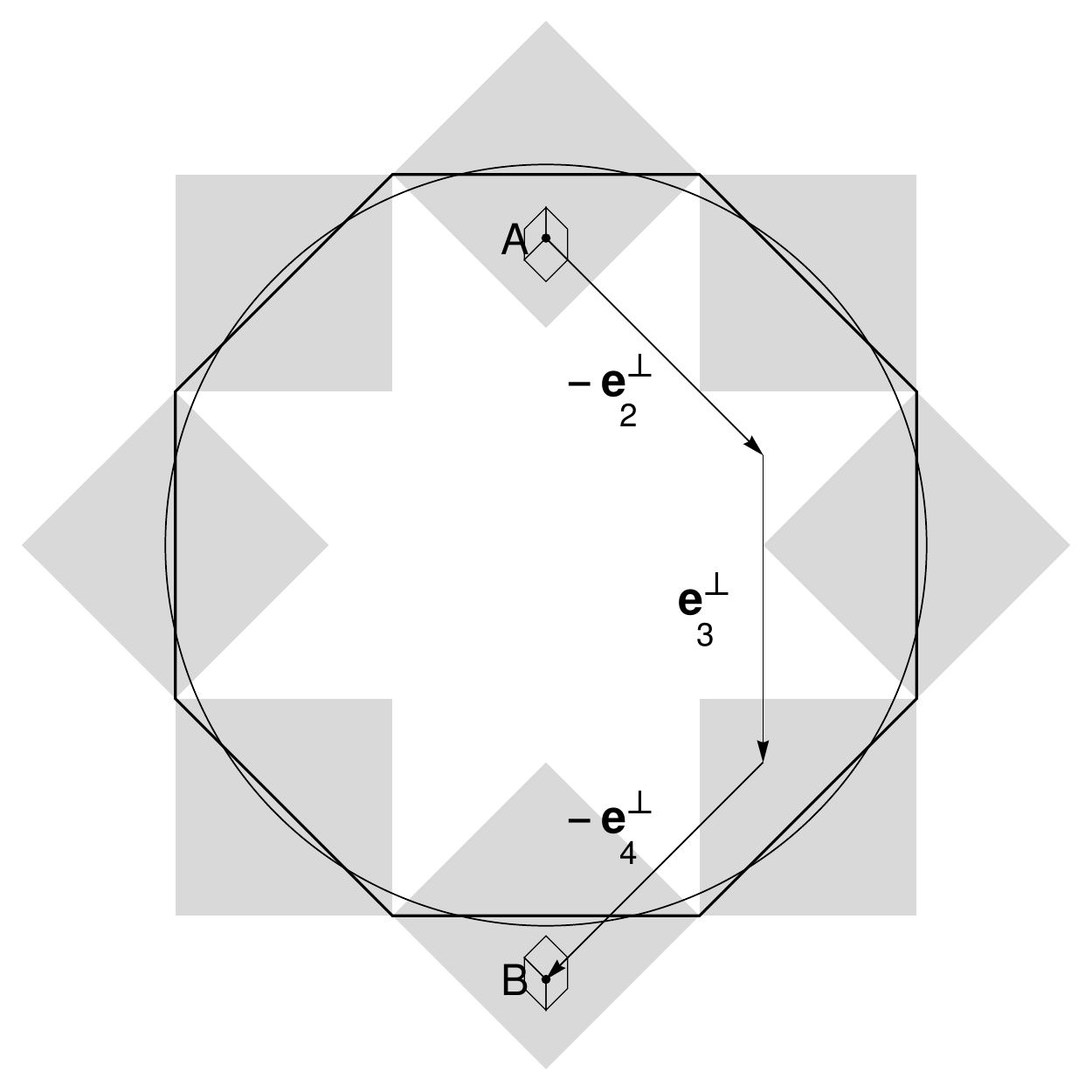} 
\caption{A pair of twin sites in real (left) and in perpendicular space (right). In each case, the three lattice displacement vectors connecting the members of the pair are shown. Grey areas represent the total zones of existence of the two sites.}
\label{phasonpair.fig}
\end{figure}

In sum, the difference in the shape of the selection windows leads to the appearance of inhomogeneities: as compared to the AB tiling, the OT has larger density fluctuations. The local intensity for the members of a pair of twin sites are slightly different, because their distance from the origin in perpendicular space are slightly different, as shown in Fig.\ref{phasonpair.fig}. Similarly, it is easy to show that the ``vacant" site inside, say, an empty hexagon has a local intensity that lies just below the cut-off value.

 Small local fluctuations could lead to depopulating the twin sites on the one hand, and populating the empty sites, on the other. This suggests that a way to remove the observed defects in the OT would be to turn on small short range repulsive interactions between atoms. One can, indeed, tune the local interaction strengths between atoms in cold gases using Feshbach resonances \cite{bloch}. In the standard Hubbard model for fermions, the interaction energy is  nonzero when the two fermions are on a single site. In the extended Hubbard model, the interaction terms concern particles on nearest neighbor sites. Both kinds of terms can be simulated in cold atom gases. By varying the relative strength of the hopping and onsite interaction  terms, it is possible, for example, to induce a superfluid to Mott insulator transition in a bosonic system \cite{jaksch}. Extensions of the Hubbard model to further near neighbors are reviewed in \cite{dutta2015}. 

Adding a small short range repulsive interaction energy will make it unfavorable to occupy simultaneously both sites of the twin pairs. One of the atoms corresponds to a larger distance from the origin in perpendicular space, and therefore has a slightly higher onsite energy than its twin. It will therefore move to an unoccupied site with similar onsite energy, corresponding to one of the empty polygons. Let us make the reasonable assumption that all pairwise nearest neighbor interactions are negligible, except for those between the atoms on twin-sites. Let the strength of the repulsive energy for a pair of twin sites be $U'$. Increasing $U'$ will make it increasingly unfavorable to occupy simultaneously both sites of a twin pair. The $U'$ value needed to shift an atom to an empty site can be estimated by the following argument. Consider a pair of twin sites, whose positions in the perpendicular space are known, for example as in Fig.\ref{phasonpair.fig}. We now consider the change in the energy of the system $\Delta E$ if one of the twin sites is vacated by the atom, which migrates to a defect "vacant" site. In this move, an atom is removed from the site whose coordinate $\vec{r}'_1$ lies within the cutoff radius but outside the octagonal window, and it is replaced elsewhere, on a site whose whose coordinate $\vec{r}'_2$ lies outside the cutoff radius but within the octagonal window. Such a move costs potential energy, and the change can be calculated by using the projected form of the potential energy in perpendicular space:
$$\Delta V = -16V_0\sum_j \cos(\vec{k}'_{j}.\vec{r}'_2)-\cos(\vec{k}'_j.\vec{r}'_1) \approx 8V_0k^2(\rho_2^2-\rho_1^2)$$
where $\rho_1$ and  $\rho_2$ are the radial distances of the initial and final positions of the atom in perpendicular space and where we used the quadratic approximation of the potential. One can calculate the maximum value of this difference, by taking $\rho_1$ to be on the midpoint of an edge of the octagonal window, and  $\rho_2$ on a vertex. The difference of distances can then be readily calculated, as a function of $p$. One thus finds that $\frac{\Delta V}{V_{max}} < \frac{\pi^2}{16}(2-\sqrt{2}) \alpha^{-2p}$. In order for such a hop to occur, the gain in onsite potential energy has to be compensated by the reduction of energy due to the destruction of a twin pair, namely, $U'$. This argument provides an upper bound to the value of the repulsive interaction needed to eliminate twin pairs. For $p=1$, for example, one gets $  U'/V_{max} \sim \Delta V/V_{max}\leq 0.06$, and the values for larger $p$ are even smaller. This estimate shows that the repulsive interaction needed would be quite small relative to the onsite potential energy term.

To transform the OT into the AB tiling, in sum, one must introduce a small additional nearest neighbor interactions, over and above the optical potential. As we noted, in cold atom systems, such repulsive interactions can be introduced, and experimentally controlled. The experimental set-up with the laser potential and repulsive interactions appears to be a feasible proposition for obtaining a perfect defect-free Ammann Beenker tiling, at least in principle. In practice, there will be problems associated with slow dynamics, trapping in metastable configurations, and thus possibly a certain amount of disorder at finite temperatures.  

The dynamics of this system can be described, to good approximation, by a tight-binding model, using the basis set of the Wannier states defined on the local minima of the optical lattice.  As we have already pointed out, the values of the diagonal terms  (the onsite energies), depend on the local environment (coordination number). The degree of localization of the Wannier functions depends on the depth of the potential minimum, and also varies to a limited extent. As for the hopping amplitude between two neighboring sites, they depend on several factors: on the distance between the sites, on the profile of the potential, and on the degree of localization of the local Wannier orbitals. These amplitudes will therefore have a variation of values for different pairs of sites. To evaluate the size of the spread of values, a detailed calculation is necessary. However, it can be argued as in \cite{epjb2014} that the hopping amplitudes are expected to fall off quickly with the distances and that the two important processes correspond to hopping along edges and across the small diagonal of the rhombus. 

With the experimental and theoretical caveats mentioned above, we see that it is theoretically possible to simulate tight-binding Hamiltonians for fermions and bosons on a perfect octagonal tiling. This quasiperiodic tiling has been a subject of theoretical investigation since the mid-eighties. To cite only a few representative works, spectral properties and quantum dynamics in tight binding Hamiltonians have been investigated by many authors using a variety of techniques (\cite{sire1,ben,ben2,moss2,trambly} . Many body effects in quasicrystals are also a question of current interest. The effect of Hubbard interactions  \cite{hubbard1997} and more recently, the fate of a local magnetic impurity  \cite{andrade2015} in this tiling were examined, motivated by recent experimental work in heavy fermion quasicrystal compounds \cite{watanuki,deguchi}. It should be possible also to simulate and study the 2D quasiperiodic antiferromagnetic Heisenberg spin model, for which theoretical works have predicted novel ground state properties \cite{wessel2003}. Effects of disorder, magnetic fields etc could also be systematically studied by means of a cold atom simulation.  

\section{Conclusions}

We have outlined a method of obtaining a two dimensional quasicrystal namely, the Ammann Beenker tiling, by trapping cold atoms in a laser generated potential. This would allow, for the first time, experimental studies of important theoretical paradigms for quasicrystals. The study of dynamics of fermions or bosons in a perfect 2D quasicrystal, and described by a Hubbard model, is one example. Another example is the experimental realization of the Heisenberg spin model. Disorder and effects due to magnetic perturbations could be investigated under controlled conditions. Progress in these problems would significantly advance our understanding of the thermodynamic and transport phenomena in real quasicrystals.


\begin{thebibliography}{20}

\bibitem{schecht}
Gratias~D. D.~Shechtman~D., Blech~I. and Cahn J.W.
\newblock {\em Phys. Rev. Let}, 53:1951, 1984.


\bibitem{europhys2013}
Anuradha Jagannathan and Michel Duneau.
\newblock {\em Europhys. Lett.}, 104:66003, 2013.

\bibitem{epjb2014}
Anuradha Jagannathan and Michel Duneau.
\newblock {\em Eur. Phys.J. B 87} 149, 2014.

\bibitem{beenker}
F.P.M. Beenker.
\newblock {\em Algebraic theory of non periodic tilings of the plane by two
  simple building blocks: a square and a rhombus,TH Report 82-WSK-04}.
\newblock Technische Hogeschool, Eindhoven, 1982.

\bibitem{bloch}
Immanuel Bloch, Jean Dalibard, and Wilhelm Zwerger.
\newblock Many-body physics with ultracold gases.
\newblock {\em Rev. Mod. Phys.}, 80:885, 2008.

\bibitem{grimm2000}
 Rudolf Grimm, Matthias Weidem\"uller  and Yuri Ovchinnikov
Adv. At.Mol.Phys. {\bf 42} 95 (2000)

\bibitem{octagonal2}
Ch. Janot and J.~M. Dubois.
\newblock {\em Quasicrystalline Materials}.
\newblock Singapore, 1988.

\bibitem{bohr}
H.~Bohr.
\newblock {\em Almost-periodic functions}.
\newblock New York, 1947.

\bibitem{besic}
A.S. Besicovitch.
\newblock {\em Almost periodic functions}.
\newblock Dover, Cambridge, 1954.

\bibitem{dunkatz}
Michel Duneau and Andr\'e Katz
Phys. Rev. Lett. {\bf 54} 2688, 1985


\bibitem{diffrac}
F.~Hippert and D.~Gratias, editors.
\newblock {\em Quasicrystals}.
\newblock Les Ulis, 1994.

\bibitem{grimmbaake}
U.~Grimm and M.~Baake.
\newblock {\em Aperiodic Order}, volume~1.
\newblock Cambridge University Press, Cambridge, 2013.


\bibitem{jaksch}
Jaksch, D. and Bruder, C. and Cirac, J. I. and Gardiner, C. W. and Zoller, P.
Phys. Rev. Lett., {\bf 81}, 3108 (1998)

\bibitem{dutta2015}
Omjyoti Dutta et el
Rep.Prog.Phys. {\bf 78} 066001, 2015
\bibitem{sire1}
C.~Sire and J.~Bellissard.
\newblock {\em Europhys. Lett.}, 11:439, 1990.

\bibitem{ben}
Vincenzo~G. Benza and Cl\'ement Sire.
\newblock {\em Phys. Rev. B}, 44:10343, 1991.

\bibitem{ben2}
B.~Passaro, C.Sire, and V.G. Benza.
\newblock {\em Phys. Rev. B}, 46:13751, 1992.


\bibitem{moss2}
J.X.Zhong and R.~Mosseri.
\newblock {\em J. Phys. I (France)}, 4:1513, 1994.


\bibitem{trambly}
G.~Trambly de~Lassardiere, C.~Oguey, and D.~Mayou.
\newblock {\em Phil. Mag.}, 91:2778, 2011.

\bibitem{hubbard1997}
A.~Jagannathan and H.~J. Schulz.
\newblock {\em Phys. Rev. B}, 55:8045, 1997.


\bibitem{andrade2015}
Andrade, Eric C. and Jagannathan, Anuradha and Miranda, Eduardo and Vojta, Matthias and Dobrosavljevi\ifmmode \acute{c}\else \'{c}\fi{}, Vladimir,
Phys. Rev. Lett. {\bf 115}, 036403 (2015)

\bibitem{watanuki}
Watanuki, Tetsu and Kashimoto, Shiro and Kawana, Daichi and Yamazaki, Teruo and Machida, Akihiko and Tanaka, Yukinori and Sato, Taku J.
Phys. Rev. B, {\bf 86}, 094201 (2012)

\bibitem{deguchi}
 Kazuhiko Deguchi,	Shuya Matsukawa, Noriaki K. Sato, Taisuke Hattori,	Kenji Ishida, Hiroyuki Takakura, Tsutomu Ishimasa	
Nat. Mater. {\bf 11}, 1013 (2012)

\bibitem{wessel2003}
Wessel, Stefan and Jagannathan, Anuradha and Haas, Stephan
Phys. Rev. Lett., { \bf 90}, 177205 (2003)




\end{thebibliography}

\end{document}